\documentstyle[epsfig]{l-aa}

\begin{document}

\def\ltsima{$\; \buildrel < \over \sim \;$}
\def\simlt{\lower.5ex\hbox{\ltsima}}

   \thesaurus{13.25.2 -- 11.19.1 -- 10.09.1}
   \title{One more surprise from the Circinus Galaxy: BeppoSAX discovery of
a transmission component in hard X-rays} 

   \author{G. Matt
          \inst{1}
   \and M. Guainazzi
          \inst{2}
   \and R. Maiolino
          \inst{3}
   \and S. Molendi
          \inst{4}
   \and G.C. Perola
          \inst{1}
   \and L.A. Antonelli
	  \inst{5,6}
   \and L. Bassani
          \inst{7}
   \and W.N. Brandt
          \inst{8}
   \and A.C. Fabian
          \inst{9}
   \and F. Fiore
          \inst{5,6}
   \and K. Iwasawa
          \inst{9}
   \and G. Malaguti
          \inst{7}
   \and A. Marconi
          \inst{3}
   \and J. Poutanen
          \inst{10}
          }


   \institute{Dipartimento di Fisica ``E. Amaldi",
              Universit\`a degli Studi ``Roma Tre",
              Via della Vasca Navale 84, I--00146 Roma, Italy
   \and Astrophysics Division, Space Science Department of
	ESA, ESTEC, Postbus 299, NL-2200 AG Noordwijk, The Netherlands 
    \and Osservatorio Astrofisico
             di Arcetri, L.~E.Fermi 5, I--50125 Firenze, Italy
   \and Istituto di Fisica Cosmica e Tecnologie Relative, Via Bassini 15,
        I--20133 Milano, Italy
      \and Osservatorio Astronomico di Roma, Via dell'Osservatorio,
        I--00044 Monteporzio-Catone, Italy
   \and SAX/SDC Nuova Telespazio, Via Corcolle 19,  I--00131 Roma, Italy
   \and Istituto Tecnologie e Studio Radiazioni Extraterrestri, CNR, Via
              Gobetti 101, I--40129 Bologna, Italy
   \and Department of Astronomy and Astrophysics, The Pennsylvania State
	University, 
	University Park, PA 16802, U.S.A.
     \and Institute of Astronomy,
        University of Cambridge, Madingley Road, Cambridge CB3 0HA, U.K.
     \and Stockholm Observatory,  SE-133 36 Saltsj\"obaden,  Sweden 
}

   \date{Received / Accepted }

   \maketitle

\markboth{G. Matt et al.}{BeppoSAX observation of the Circinus Galaxy}

   \begin{abstract}
The Circinus Galaxy has been observed, for the first time above 10 keV, 
by BeppoSAX. An excess emission above the extrapolation of the 
best fit 0.1--10 keV spectrum is apparent in the PDS data. The
most likely explanation 
is that we are observing the nucleus through a
column density of $\sim4\times10^{24}$ cm$^{-2}$, i.e. 
large enough to completely block the radiation below 10 keV but small enough
to permit partial transmission above this energy.
   \end{abstract}

%

\section{Introduction}

The Circinus Galaxy hosts one of the nearest Active Nuclei
($d\sim$4 Mpc) and has been therefore much studied after its discovery (Freeman 
et al. 1977), despite the
low Galactic latitude and therefore the relatively large interstellar 
absorption (N$_{\rm H, Gal}\sim 3 \times 10^{21}$ cm$^{-2}$). 

The intense coronal lines (Oliva et al. 1994, Moorwood et al. 1996), the 
[O {\sc iii}] ionization cone (Marconi et al. 1994), the 
variable H$_2$ maser emission (Greenhill et al. 1997), the broad H$\alpha$
line in polarized light (Oliva et al. 1998a), and the reflection--dominated
2--10 keV X--ray spectrum (Matt et al. 1996, hereinafter M96)
support the Seyfert 2 classification for the Circinus Galaxy. 
The integrated IR luminosity is about 4$\times10^{43}$ erg s$^{-1}$,
a large fraction of it being probably reprocessing of the nuclear radiation
(Moorwood et al. 1996; Maiolino et al. 1998). 

The ASCA observation (M96) revealed a flat X--ray spectrum with
several emission lines, including a very prominent iron K$\alpha$ line at
about 6.4 keV. These findings were interpreted by M96 in terms of reflection
by cold matter (possibly the inner wall of the putative torus envisaged
by unification models, e.g. Antonucci 1993) 
of an otherwise invisible nuclear emission. The absence of any transmitted
component in the ASCA band (i.e. up to 10 keV) indicates a column density
for the absorbing matter in excess of 10$^{24}$ cm$^{-2}$, i.e. a 
Compton--thick source. 

The Circinus Galaxy was observed by BeppoSAX (Boella et al. 1997)
on 1998 March 24
as part of a Core Program aiming
to study the broad band spectra of Compton--thick Seyfert 2 galaxies. 
Here we report on the spectral analysis of the source, with particular
emphasis on the high energy part of the spectrum. 

\section{Observation and data reduction}

We will discuss here data from three
instruments: the LECS, the MECS and the PDS.
The MECS is presently composed of two units (after the failure of a third
one), working in the 1--10 keV energy range. At 6~keV,
the energy resolution is  $\sim$8\%  and the angular resolution is
$\sim$0.7~arcmin (FWHM). The LECS has characteristics similar to
that of the MECS in the overlapping band, but its energy band
extends down to 0.1 keV. The PDS is a passively collimated detector
(about 1.5$\times$1.5 degrees f.o.v.) working in the 13--200 keV energy range.
The effective exposure time of the observation was
1.37$\times$10$^5$ sec (MECS), 8.37$\times$10$^4$ sec (LECS)
and 6.32$\times$10$^4$ sec (PDS). 

A second source is clearly visible in the LECS and MECS images 
at about 5' from the Circinus Galaxy.
 This source was discovered by ASCA (M96), and is also
present in  the ROSAT HRI image (Ward 1998). 
The nominal positions of both sources, as measured by BeppoSAX, 
are shifted by the same amount (about 2') and in the same direction 
with respect to the HRI 
positions (which in turn are consistent with the optical position, for
Circinus, and the ASCA position, for the second source). 
Only the Y and X startrackers, one gyro and
the solar sensor were in use during the observation, but not the
Z startracker, i.e. that coaligned with the telescopes.
In this configuration, absolute position reconstruction can
be in error up to 3'.
The second source is  about 3.7 times fainter in both instruments. The
spectrum may be well fitted either by a power law with $\Gamma\sim$2.35 and 
N$_{\rm H} \sim 1.3\times10^{22}$ cm$^{-2}$ or 
a plasma model with $kT\sim$6 keV and 
N$_{\rm H} \sim 3.4\times10^{21}$ cm$^{-2}$. In both cases, 
the extrapolation to the PDS band lies more than two orders
of magnitude below the data, and its contribution
to the PDS data may be neglected. Of course, in principle
the excess emission in the PDS band, described in Sec.3, could be due to
this source, if it is heavily absorbed (N$_{\rm H} \simeq 10^{24}$ cm$^{-2}$).
This would imply 
a physical scenario similar, but much more extreme and then less
plausible, to that discussed below for Circinus. 

To avoid, as far as possible, contamination from the second source we
extracted data for both the LECS and the MECS in 2' radii centred on the
source. The background subtraction was performed using
blank sky spectra from the same region of the detector
field of view. The resulting count rates are 4.41$(\pm0.08)\times$10$^{-2}$ in
the LECS and 1.050$(\pm0.009)\times$10$^{-1}$ counts/s in the MECS.
PDS data have been reduced following the standard procedures described in
Matt et al. (1997). In addition, we employed crystal temperature
dependent Rise Time thresholds, which lead to a reduction by
a factor up to 50\% of the instrumental background. 
 The PDS count rate
is 2.01$\pm$0.04 counts/s. 

The spectra were fitted using the calibration matrices
released on September 1997. In the fit procedure, the LECS normalization was
left free to vary with respect to the MECS one, while the PDS normalization
was fixed to 0.8 (Grandi et al. 1997). The residual
uncertainty on the PDS absolute flux calibration ($\simeq \pm 10\%$)
does not affect substantially the following results. 
All quoted fluxes refer to the MECS. 
Errors correspond to 90\% confidence level for two
interesting parameters (i.e. $\Delta\chi^2$=4.61). 

Meaningful LECS and MECS light curves of the Circinus Galaxy cannot be
easily constructed:
due to the problems in the aspect reconstruction, a small extraction radius
results in a fictitious variation of the count rate due to the 
wobbling; if a large extraction radius is instead chosen, the contamination
from the second source is significant. In the latter case, a variation up
to 10\% on time scales of hours is observed in the MECS. 
If this is fully attributed
to the second source, as should be if the interpretation of the Circinus
spectrum  below 10 keV as pure reflection component is correct, this would
imply a $\sim$50\% amplitude variations in this source. Finally, the PDS
light curve is consistent with being constant (with a $\sim$40\% 
upper limit to the variability amplitude). 

\section{Spectral analysis and results}


We firstly analysed the spectrum below 10 keV, to verify the consistency
with the ASCA results (M96). The spectrum is very flat with a huge iron 
K$\alpha$ line, as 
observed by ASCA and interpreted by M96 as reflection dominated emission,
the reflection probably occurring in circumnuclear matter (hereinafter 
identified, for simplicity, with the ``torus" familiar from the
unification models). 
The 2--10 keV flux is 1.4$\times10^{-11}$ erg cm$^{-2}$
s$^{-1}$, consistent with the ASCA one.  
At least other five lines,
to be identified with Si, S, Ar and Ni K$\alpha$ 
and Fe  K$\beta$ lines, are required, as well as excess emission below 
a few keV. 
The properties of the emission lines, as well as of the soft excess,
are discussed in detail elsewhere (Guainazzi et al., 1998).



\begin{figure}
\epsfig{file=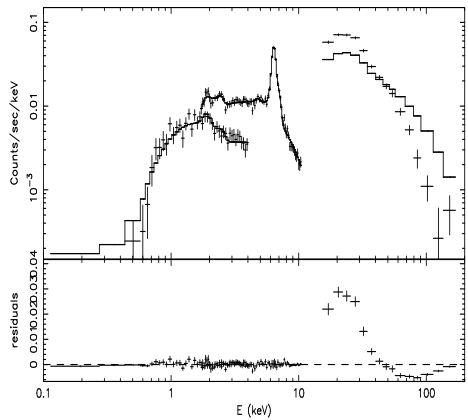, height=7.cm, width=7.cm, angle=-90}
\caption{LECS+MECS best fit model extrapolated to the PDS data. The model 
is composed by a pure reflection component ($\Gamma_h$=1.60), a soft
power law ($\Gamma_s$=1.45) plue the emission lines.}
\label{leme_extr}
\end{figure}

When the model described above is extrapolated
to the PDS band, a huge excess emission is apparent (see Fig.\ref{leme_extr}).
No known bright sources are present in the field of view 
of the PDS. The Galactic longitude 
(311$^{\circ}$) and latitude (-3.8$^{\circ}$)  of 
the source are about 10$^{\circ}$
outside the bulk of the Galactic Ridge emission (Warwick et al. 1985). 
Following the modelling of Yamasaki et al. (1997), we found that the 
contribution of the Galactic Ridge to the PDS count rate is at most 7\%.
Therefore, it is very likely that 
we are observing a further component in the Circinus spectrum, possibly
the nuclear radiation piercing throughout a very large column density absorber, 
as in the case of NGC~4945 (Iwasawa et al. 1993; Done et al. 
1996) and Mkn~3 (Cappi et al. 1998)
In order to conceal itself  below 10 keV, the absorber 
column density must exceed 10$^{24}$ cm$^{-2}$, i.e. must be Compton--thick. 
This implies that the photons which eventually emerge
may have suffered one or more Compton scatterings. 
We have therefore adopted a model, based on MonteCarlo simulations, 
which properly treats photon transfer in very high column density matter (work
in preparation), including Compton downscattering and the Klein--Nishina 
decline. We assumed a spherical geometry 
as the simplest approximation of the (actually unknown)
distribution of the matter. If a plane--parallel geometry, with normal
illumination, is employed, the results do not change significantly. 

We have then fitted the whole spectrum with the following model:

\begin{eqnarray}
F(E) & = & [ A_1 Tr(E,N_{\rm H,1},\Gamma_h,E_{c}) 
+  A_2 R(E,\Gamma_h,E_{c})  \cr
~ & ~ & + A_3  E^{-\Gamma_s} + {\sl emission~lines}]  
e^{-N_{\rm H,2}\sigma_{\rm ph}} 
\end{eqnarray}

\noindent
where 

\begin{equation}
Tr = \int^{\infty}_{E} G(E,E_0, N_{\rm H,1}) E_0^{-\Gamma_h}e^{(-E_0/E_c)} dE_0 
\end{equation}

\noindent
and $G(E,E_0)$ is the Green function relating input and transmitted
spectra.
$\Gamma_h$ and $\Gamma_s$ are the power law indices of the
hard (nuclear) component and of the soft component, respectively; 
$N_{\rm H,1}$ is the column density of the torus and $N_{\rm H,2}$
that of any external matter along the line  of sight. $R(E,\Gamma_h,E_{c})$
is the pure reflection spectrum, obtained using 
{\sc pexrav} in XSPEC fixing cos($\theta$) to 0.45 and 
{\verb!rel_refl!} to --1. 

\begin{figure}
\epsfig{file=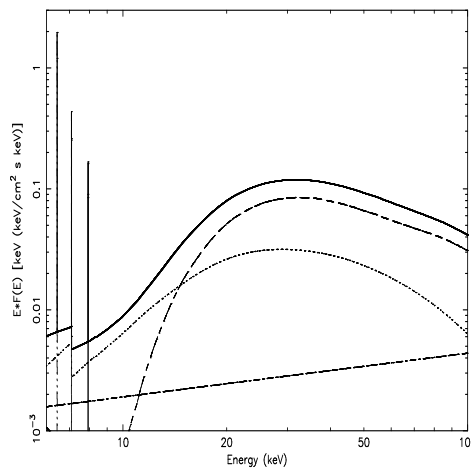, height=7.cm, width=7.cm, angle=-90}
\caption{Best fit reflection + transmission model (in $\nu F_{\nu}$).}
\label{trans_model}
\end{figure}

The 20--100 (50--150) keV total flux is 2.3$\times10^{-10}$ 
(0.9$\times10^{-10}$) erg
cm$^{-2}$ s$^{-1}$, which makes the Circinus Galaxy one of the brightest 
AGN in hard X--rays.
The best fit parameters are summarized in Table 1 (model 1), while the 
best fit model is shown in Fig.~\ref{trans_model}.
$\Gamma_s$ and $\Gamma_h$ are very similar, and consistent 
within the errors. It is possible that the soft emission is due to
scattering of the nuclear radiation by ionized matter (Guainazzi et al. 1998). 

The normalization of the nuclear component, $A_1$, is about 5 times
larger than that of the reflection component,  $A_2$; therefore, 
the solid angle subtended by the reflecting matter to the nucleus
{\it and} visible at infinity is $\sim$0.2$\times2\pi$.
Assuming a torus geometry like that adopted by Ghisellini et al. (1994)
(i.e. a conical geometry with half--opening angle of 30$^{\circ}$), 
we derive an inclination angle $\simlt$40$^{\circ}$. 

The 2--10 keV nuclear luminosity, after correction for absorption,
ranges from 3.4$\times10^{41}$ to 1.7$\times10^{42}$ erg s$^{-1}$ 
(assuming a distance of 4 Mpc). 
The IR luminosity (as defined by Mulchaey et al. 1994)
is 20 to 100 times larger. 
This IR/(2--10 keV) ratio is somewhat higher than the average value for
Seyfert galaxies, but by no means unique
(see Fig. 2a of Mulchaey et al. 1994; Awaki 1997). If a significant
fraction of the IR emission is
reprocessing of the nuclear component, a large
UV nuclear emission is implied, qualitatively in agreement with the
calculations of Moorwood et al. (1996) and Oliva et al. (1998b) 
based on the IR lines.
The X-ray to OIII ratio is typical for Seyfert galaxies (Bassani et al. 1998). 

\begin{table}
\caption{Best fit parameters. See text for details on the models. }
\vspace{0.05in}
\begin{tabular}{ccc}
\hline
\hline
~ & 1 & 2 \cr
\hline
~ & ~ & ~\cr
N$_{\rm H,1}$~(10$^{24}$~cm$^{-2}$) & 4.3$^{+0.4}_{-0.7}$ & 2.9$^{+0.6}_{-0.7}$ \cr
A$_1$ & 0.11$^{+0.07}_{-0.08}$  & 0.12$^{+0.15}_{-0.08}$ \cr
$\Gamma_h$ & 1.56$^{+0.16}_{-0.51}$ & 1.62$^{+0.21}_{-0.43}$ \cr
E$_{\rm C}$~(keV) & 56$^{+8}_{-23}$ & 81$^{+37}_{-37}$ \cr
A$_2$ & 2.02($^{+0.65}_{-0.98})10^{-2}$ & 2.10($^{+0.68}_{-0.95})10^{-2}$ \cr
$\Gamma_s$ & 1.64$^{+0.13}_{-0.43}$  & 1.60$^{+0.18}_{-0.41}$ \cr
A$_3$ & 8.3($^{+1.2}_{-1.7})10^{-4}$ & 8.0($^{+1.4}_{-1.7})10^{-4}$ \cr
N$_{\rm H,2}$~(10$^{21}$~cm$^{-2}$)  & 3.6$^{+1.0}_{-1.0}$ & 3.5$^{+1.1}_{-1.3}$ \cr
$\chi^2$/d.o.f. & 122/121 & 119/121 \cr
~ & ~ & ~\cr
\hline
\hline
\end{tabular}
\end{table}

We have also tested a different scenario (model 2 in Table~1), 
in which the transmitted component is simply a fraction of 
the total reflection component, i.e. the reflecting matter 
is partially covered. 
This case would represent a scenario in which the line of sight 
towards the nucleus is completely blocked (i.e $N_{\rm H}$ in excess
of $\sim$10$^{25}$ cm$^{-2}$), but the optical depth diminishes at
higher torus latitudes, so permitting transmission of 
some of the radiation reflected by the far side of the torus. The spectrum 
is similar to that of eq.~1,  but substituting, in the integral of eq.~2, 
the reflection component to the power law one. 
From the statistical point of view, this model is as good as model 1.
The $A_1$/$A_2$ ratio is now the ratio between the solid angles subtended
by the covered and uncovered matter to the source and visible at infinity.
From Table~1, it can be seen that the covered fraction of the reflector is
5 times larger than the uncovered one. No estimate of the 
nuclear luminosity is possible in this model, and therefore no information
on the torus geometry can be derived.

The observed value of the torus column density is 4.3$\times10^{24}$ cm$^{-2}$
in model 1 (and not much lower in model 2), i.e.
about 350 times larger than the value expected from the extinction in the V
band (A$_V$=5 mags)
measured from narrow lines Balmer decrement (Oliva et al. 1994)
assuming a Galactic dust--to--gas ratio. Very likely, the absorbers responsible
for the obscuration of the nucleus and the
narrow line region are different, the former being distributed on much
smaller scales. Granato et al. (1997) fitted the ISO spectrum
(Moorwood et al. 1996) with their torus model, and found  A$_V$=50 mags for the
torus, corresponding to a $N_{\rm H}$ 
more than one order of magnitude lower than our measured value, thus implying
a dust depleted material. If their estimate is correct,
the absorbing matter should then  
be located inside the dust sublimation radius which, given the nuclear
luminosity derived from model 1, is of order of 0.01 pc,
very close indeed to the central engine and suggestive of a possible
identification of the absorbing material with the BLR.

Finally, a high energy cutoff with $e$-folding energy below 100 keV 
is required by the data in both models. 
This, together with the rather flat intrinsic power law, make
Circinus similar, as far as the primary 
spectral shape is concerned, to NGC~4151 (see e.g. Piro et al. 1998).

\section{Discussion}

The main results of the BeppoSAX observation of the Circinus Galaxy can be
summarized as follows:

\medskip
\noindent
{\sc i)} The spectrum below 10 keV is dominated by a reflection component, in
agreement with the ASCA result (M96). 

\medskip
\noindent
{\sc ii)} The extrapolation of the LECS+MECS best fit 
spectrum falls short of the
PDS data. A further component transmitted through a $\sim$4$\times10^{24}$ 
cm$^{-2}$ absorber is required, but it is not
possible to establish, on statistical ground, 
whether we are looking at nuclear radiation or at
a reflection component passing through the outer edge of the torus, even
if the rather {\it ad hoc} geometry required in the 
second scenario makes the first hypothesis more appealing. To further address
this issue, one has to search for
short term variability, a task which requires much more sensitive hard
X--ray detectors, like those on board Constellation-X.
If the transmitted radiation is actually the nuclear one, by comparing the
nuclear luminosity and that of the reflection component
the inclination angle of the
system is constrained to be less than 40$^{\circ}$, 
if the torus geometry proposed by Ghisellini et al. (1994) is adopted.

Finally, it is worth noting that two of the nearest AGNs, the Circinus
Galaxy and NGC~4945, both have absorbing matter with $N_{\rm H}$
between 10$^{24}$ and 10$^{25}$ cm$^{-2}$, suggesting that this kind 
of source is rather common in the Universe. 
This is particularly interesting because with such an absorber
most of the energy in the X--ray band comes out at a few tens of keV (Fig.~2),
where the energy density of the Cosmic X--ray Background peaks.

\begin{acknowledgements}
We acknowledge the BeppoSAX SDC team for providing pre--processed event files
and the support in data reduction.  
The following institutions are acknowledged for financial support:
Italian Space Agency (GM, GCP, RM, SM, FF, AM), ESA (MG, research fellowship), NASA 
(WNB, LTSA program), Royal Society (ACF), PPARC (KI).
\end{acknowledgements}

\end{document}